\documentclass[manuscript]{acmart}

\DeclareUnicodeCharacter{2212}{-}

\AtBeginDocument{%
  \providecommand\BibTeX{{%
    \normalfont B\kern-0.5em{\scshape i\kern-0.25em b}\kern-0.8em\TeX}}}

\acmConference[--]{--}{2024}{--}
\acmISBN{}

\begin{document}

\title[Ironies of Generative AI]{Ironies of Generative AI: Understanding and mitigating productivity loss in human-AI interactions}

\author{Auste Simkute}
\authornote{Both authors contributed equally to this research.}
\affiliation{%
  \institution{University of Edinburgh}
  \city{Edinburgh}
  \country{United Kingdom}}
\email{a.simkute@sms.ed.ac.uk}

\author{Lev Tankelevitch}
\authornotemark[1]
\email{lev.tankelevitch@microsoft.com}
\affiliation{%
  \institution{Microsoft Research}
  \city{Cambridge}
  \country{United Kingdom}
}

\author{Viktor Kewenig}
\email{ucjuvnk@ucl.ac.uk}
\affiliation{%
  \institution{University College London}
  \city{London}
  \country{United Kingdom}
}

\author{Ava Elizabeth Scott}
\affiliation{%
  \institution{University College London}
  \city{London}
  \country{United Kingdom}
}
\email{ava.scott.20@ucl.ac.uk}

\author{Abigail Sellen}
\affiliation{%
  \institution{Microsoft Research}
  \city{Cambridge}
  \country{United Kingdom}}
\email{asellen@microsoft.com}

\author{Sean Rintel}
\affiliation{%
  \institution{Microsoft Research}
  \city{Cambridge}
  \country{United Kingdom}}
\email{serintel@microsoft.com}

\renewcommand{\shortauthors}{Simkute and Tankelevitch, et al.}

\begin{abstract}
Generative AI (GenAI) systems offer opportunities to increase user productivity in many tasks, such as programming and writing. However, while they boost productivity in some studies, many others show that users are working ineffectively with GenAI systems and losing productivity. Despite the apparent novelty of these usability challenges, these `ironies of automation' have been observed for over three decades in Human Factors research on the introduction of automation in domains such as aviation, automated driving, and intelligence. We draw on this extensive research alongside recent GenAI user studies to outline four key reasons for productivity loss with GenAI systems: a shift in users' roles from production to evaluation, unhelpful restructuring of workflows, interruptions, and a tendency for automation to make easy tasks easier and hard tasks harder. We then suggest how Human Factors research can also inform GenAI system design to mitigate productivity loss by using approaches such as continuous feedback, system personalization, ecological interface design, task stabilization, and clear task allocation. Thus, we ground developments in GenAI system usability in decades of Human Factors research, ensuring that the design of human-AI interactions in this rapidly moving field learns from history instead of repeating it.
\end{abstract}

\ccsdesc[500]{Human-centered computing~HCI theory, concepts and models}
\ccsdesc[500]{Computing methodologies~Artificial intelligence}
\ccsdesc[500]{Human-centered computing~Interaction design process and methods}

\keywords{Generative AI, Copilot, Large Language Models, Human Factors, human-automation interaction, human-centered design, human-AI interaction, usability}

\received{20 February 2007}
\received[revised]{12 March 2009}
\received[accepted]{5 June 2009}

\maketitle

\section{Introduction}\label{sec:intro}

Generative artificial intelligence (GenAI) systems, such as large language models (LLMs) that can generate novel content and perform many other tasks, present myriad opportunities and challenges to humans in knowledge-intensive domains. GenAI applications have emerged in domains such as healthcare \cite{nova_generative_2023}, research \cite{lund_chatting_2023}, writing \cite{dang_choice_2023,chen_large_2023}, creative work \cite{parra_pennefather_ai_2023,oppenlaender_creativity_2022,gmeiner_exploring_2023,kulkarni_word_2023}, consulting \cite{dellacqua_navigating_2023}, and recruitment \cite{budhwar_human_2023}. Software engineering has been particularly impacted, with GenAI-assisted programming tools, such as GitHub Copilot \cite{friedman_introducing_2021}, being increasingly used to support software engineering practices and perform tasks such as auto-completing code \cite{kim_code_2021}, translating code across languages, and answering programming questions, among others \cite{ross_programmers_2023, sarkar_what_2022}.

GenAI’s ability to solve domain-specific problems speaks to its potential to augment human performance and transform productivity. Recent research already suggests the enormous positive impact these systems could have on workers' performance in domains including programming \cite{peng_impact_2023}, writing \cite{noy_experimental_2023}, law \cite{choi_ai_2023}, and consulting \cite{dellacqua_navigating_2023}. Based on this research, the expectation is that new tools will often free up users’ time and allow them to focus on higher-level tasks, increasing their productivity. However, when using the new tools in practice, many users, such as programmers, report increased cognitive load, frustration, and time spent on the tasks that GenAI is intended to support. Feedback from Copilot users, as well as usability studies of GenAI-driven programming tools, suggest that, in some cases, using GenAI support can, in fact, lead to productivity loss. For example, software engineers and novice programmers struggle to effectively prompt systems, debug generated code, lose their state of flow when interrupted by long code suggestions, and get stuck in ineffective practices, such as reviewing, editing and then ultimately deleting suggestions \cite{prather_its_2023, barke_grounded_2023, sarkar_what_2022}. Similar observations are emerging in creative domains, where graphic \cite{oppenlaender_creativity_2022,kulkarni_word_2023} and manufacturing \cite{gmeiner_exploring_2023} designers struggle with prompt engineering and other aspects of GenAI interaction. %
This suggests that the potential of GenAI systems to boost productivity may not be guaranteed, evenly distributed, or fully exploited.  

These observations mirror the long line of Human Factors studies exploring human-automation interactions in safety-critical systems in aviation, industrial plants, and other areas \cite{lee_human_2009, endsley_here_2017}. Indeed, they reflect the `ironies of automation' \cite{bainbridge_ironies_1983}, which capture the idea that the more advanced an automated system is, the more important the human operator may be.\footnote{\citet{endsley_ironies_2023} makes a similar parallel between the ironies of automation and the challenges of modern AI systems; however, whereas they cover both generative and non-generative AI and take a high-level view of AI, the current paper focuses specifically on GenAI and examines concrete usability challenges documented in recent user studies of GenAI systems.} Despite automation taking over human manual control in areas where it is expected to provide superior performance, humans are still left to supervise automation. However, operators might have insufficient support to supervise, and so instead of being supported by automation, they find themselves cognitively overburdened, trying to decipher systems' outputs and spot errors. Similarly, in the context of GenAI, users’ roles have shifted from producing output to evaluating it, often with little contextual information and situational awareness. This is exacerbated by GenAI tools' ability to produce outputs at a capacity too demanding for adequate evaluation, with questionable reliability, and with poor explainability \cite{liao_ai_2023,chen_next_2023,schellaert_your_2023}. Moreover, poor system and interface design can result in unhelpful restructuring of workflows, which increases cognitive load and undermines productivity gains \cite{bainbridge_ironies_1983}. This is echoed in programmers' experiences and feedback around Copilot features \cite{sarkar_what_2022,barke_grounded_2023,prather_its_2023}, with evidence of similar effects emerging in other domains \cite{dang_choice_2023, gu_how_2023-1,gmeiner_exploring_2023}. Finally, as a result of which tasks get automated, as well as poor system design, automation often makes easy tasks easier while making hard tasks even harder. This same pattern is now %
being observed in usability studies of GenAI systems \cite{sarkar_what_2022,barke_grounded_2023}.

In this paper, we answer recent calls for bridging Human Factors and Human-Computer Interaction research to advance human augmentation by AI and human-AI interactions \cite{chignell_evolution_2023}. Extrapolating from over 30 years of Human Factors research on the `ironies' of human-automation and productivity loss, we synthesize an overview of the usability and productivity challenges observed in recent GenAI user studies. We demonstrate how these challenges emerging in GenAI systems mirror those experienced by operators when automation was introduced to their workflows decades ago. Based on these parallels, we highlight key areas of productivity loss and provide insights into the human factors leading to these issues, exploring aspects including feedback, situational awareness, cognitive workload, workflow disruptions and others. We focus primarily on programming due to the early adoption of tools like GitHub Copilot and the accompanying usability research, but we also reflect on emerging studies from other domains, such as healthcare, writing, and design, showing that these issues are not limited to a single domain. Moreover, we discuss potential design solutions, emphasizing the importance of following the Human Factors principles of feedback and flexibility when designing GenAI systems. We suggest that the fast-paced innovation of GenAI will benefit from the decades of Human Factors research in order to design GenAI systems that truly harness the full productivity potential of this technology. In summary, our paper makes the following contributions:

\begin{enumerate}
\item Based on Human Factors research and a synthesis of recent GenAI studies, we identify key challenges that can lead to productivity loss, grouped into four broad categories: (i) the production-to-evaluation shift, (ii) unhelpful workflow restructuring, (iii) task interruptions, and (iv) task-complexity polarization.

\item We provide potential design directions from Human Factors research that address each category of challenges: (i) continuous feedback, (ii) system personalization, (iii) ecological interface design, (iv) main task stabilization and timing, and (v) clear task allocation. %
Throughout, we also emphasize the importance of following the Human Factors principles of feedback and flexibility.

\item We motivate further research into the impact of GenAI systems on aspects such as situational awareness and cognitive workload to better understand systems' unintended effects on human performance. We also encourage future researchers to take advantage of the plethora of relevant Human Factors work to enrich their understanding of existing human-GenAI interaction issues and anticipate others.
\end{enumerate}

\section{Productivity Challenges of Generative AI Automation}\label{sec:challenges}
Here, we outline the key productivity challenges that have been observed in human-automation interaction over decades of Human Factors research and are now becoming apparent in user studies of GenAI systems. Our focus is on GenAI \textit{systems}, the integrated whole comprising GenAI models and interfaces. Some challenges pertain to \textit{GenAI models} (e.g., issues around prompting), and some pertain to \textit{interface design} (e.g., issues around task interruptions). 

We begin with challenges related to the shift from manual control or production to a more passive supervisory role of the user, such as monitoring and evaluation of AI outputs (Section \ref{subsec:shifteval}). We explore specific aspects related to this shift, such as reduced situational awareness, the contributory factors of automation's high capacity, complexity and opaqueness, reliability, and potential resultant complacency and over-reliance. We then outline how the introduction of automation such as GenAI can unhelpfully restructure users' workflows, stifling their productivity (Section \ref{subsec:restructuring}). We focus on how the introduction of new tasks, such as prompting or output adaptation, can affect user performance and how workflow restructuring can lead to loss of task sequence and feedback. We also explore the influence that task interruptions from AI suggestions can have on users' productivity (Section \ref{subsec:interruptions}). Finally, we explore how automation such as GenAI can paradoxically lead to easy tasks being made easier and hard tasks made harder, a phenomenon we refer to as `task-complexity polarization' (also known as "clumsy automation" in Human Factors research \cite{wiener_flight-deck_1980}; Section \ref{subsec:polarization}). Figure 1 outlines the four types of challenges.

\begin{figure}[h]
\centering
  \includegraphics[width=1\linewidth]{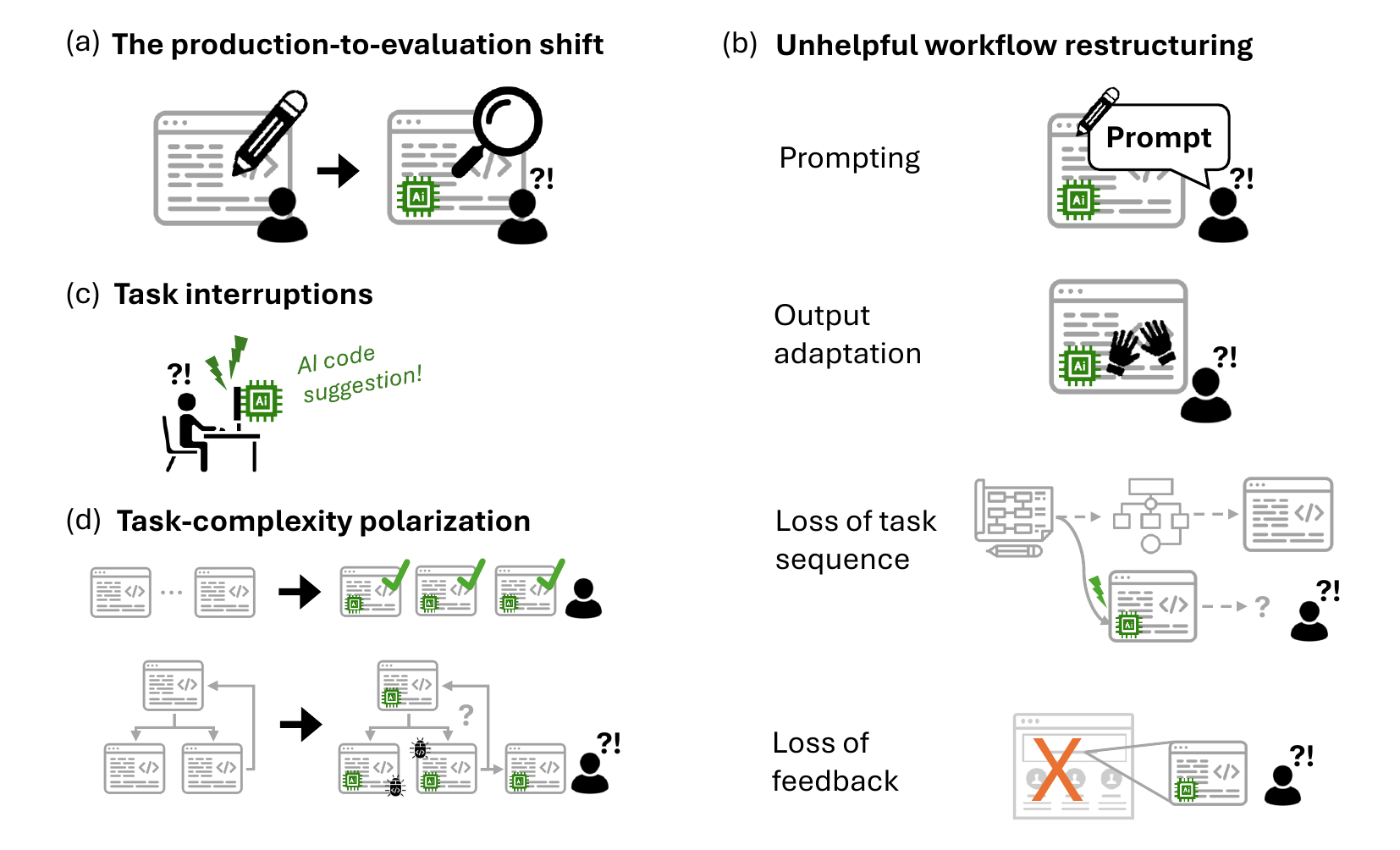}
  \caption{Productivity challenges of Generative AI automation: (a) the production-to-evaluation shift, in which users' situational awareness of their working environment is reduced, increasing the cognitive demand required to evaluate AI outputs; (b) unhelpful workflow restructuring, including the addition of new challenging tasks of prompting systems and adapting outputs, a loss of task sequence due to AI suggestions or other changes, and a loss of feedback when AI suggestions are presented without the relevant context; (c) task interruptions from automated AI suggestions; and (d) task-complexity polarization, in which automation tends to make easy tasks easier and hard tasks harder when implemented in practice.}
  \Description{The figure illustrates four key challenges of Generative AI automation using labels and illustrative icons. The first challenge is a shift from production to evaluation. The illustrative icon demonstrates an interface with a pencil (editing tool) directed with an arrow icon to another interface (GenAI) with a magnifying glass and a questioning human figure (confusing evaluation). Another challenge is unhelpful workflow restructuring. Here, four illustrative icons demonstrate different types of restructuring: prompting, shown as an interface (GenAI) with a pencil (editing tool) on a prompt icon coming into an interface; output adaptation, shown as an interface (GenAI) with two hands on the interface; loss of task sequence as an icon of two different sequences, one without AI, connected with arrows, showing a continuation of following the steps, and another is going from a task step to a GenAI interface and then a question mark; finally, loss of feedback as an icon of a GenAI output and a crossed interface demonstrating lost context (e.g., website search information). The third challenge is task interruption, demonstrated as a human icon sitting in front of a computer (GenAI) with AI suggestions written on the side and lightning bolts. The last challenge is task-complexity polarization. Using interface icons, it demonstrates a linear, simple sequence of tasks, with an arrow to the right showing three more simple tasks with green checkmarks (easy tasks made easier); it also shows a complex hierarchical structure of connected interfaces (tasks), with an arrow to the right showing a similar but more complex hierarchical structure of interfaces (tasks) with GenAI icons, software bugs, and a question mark (hard tasks made harder).}
  \label{fig:fig_challenges}
\end{figure}

\subsection{The production-to-evaluation shift}\label{subsec:shifteval}
Decades ago, the introduction of automation shifted many manual control tasks to monitoring tasks, leaving humans to supervise the automation \cite{sheridan_human_2012}. However, monitoring (or vigilance) is tedious and requires attention, and can, therefore, paradoxically impose a considerable workload on humans \cite{warm_vigilance_2008,grubb_effects_1995}. For example, when automation was introduced in the aviation context (e.g., detection of air traffic in an aircraft’s vicinity), pilots’ workload was not reduced but moved to supervising activity. Pilots reported spending more time interacting with automation and trying to understand it instead of concentrating their efforts on their primary task of flying the aircraft \cite{rudisill_line_1995}. In other domains, operators supervising automation also spent a significant amount of time and effort learning how to manage the new technology \cite{baxter_ironies_2012} (see Section \ref{subsubsec:adaptation}).

GenAI workflows have introduced a similar shift from manual control to monitoring—in this case, from the production of outputs to their evaluation—with \cite{sarkar_exploring_2023} terming this new user role “critical integration” (see Figure \ref{fig:fig_challenges}a).\footnote{This shift from production to evaluation is \textit{relative} rather than absolute, as, for example, crafting prompts still constitutes a form of production (see Section \ref{subsubsec:prompting}).} In AI-assisted coding, users spend extended periods reviewing and validating code suggestions \cite{barke_grounded_2023, vaithilingam_expectation_2022}, sometimes at the expense of other productive tasks like writing code or running tests \cite{weisz_better_2022, vaithilingam_expectation_2022}. Some programmers have said that working with Copilot felt like a “proofreading task” \cite{weisz_better_2022}. Accordingly, in some cases, working with current GenAI systems might not benefit users relative to a more manual approach. For example, when \cite{vaithilingam_expectation_2022} compared programmers' experience with Copilot versus traditional autocomplete, they found that Copilot participants failed to complete their tasks more often. When they did complete them, they were no faster than those who used autocomplete. Vaithilingam et al. \cite{vaithilingam_expectation_2022} suggest that assessing the correctness of generated code created an efficiency bottleneck, often leading participants down an unsuccessful path of debugging. This not only took time out of their main task, thereby decreasing productivity, but also required a significant amount of cognitive effort. A similar shift towards evaluation of outputs has been observed in consultancy \cite{dellacqua_navigating_2023}, and in creative writing, where most of the writing time is now being replaced by editing AI-generated text \cite{noy_experimental_2023}. Overall, practitioners from various domains, such as advertising, education, business and law, overwhelmingly agree that GenAI outputs will require supervision \cite{woodruff_how_2023}. 

\subsubsection{Reduced situational awareness}\label{subsubsec:sitawareness}
A key reason why monitoring automation (like evaluating GenAI outputs) is so demanding is that, due to processing being relatively more passive, it reduces operators’ situational awareness: their perception of data and elements of the situation, comprehension of the situation, and the projection of future status \cite{endsley_measurement_1995}. Passive processing resulting in decreased situation awareness has been observed with experienced air traffic controllers \cite{endsley_effect_1997, metzger_role_2001} and in other automated tasks \cite{manzey_human_2012}. Low situation awareness significantly decreased operators' ability to effectively monitor and observe errors in the automation and to determine whether the given situation is outside the bounds of automation capabilities \cite{jones_sources_1996}. 

Evidence suggests that users of GenAI systems similarly experience reduced situational awareness. For example, participants in \cite{vaithilingam_expectation_2022} reported that their debugging of AI-generated code was hampered because they could not use their intuition about where the bug might be and instead ended up refactoring or abandoning the code entirely. This is echoed by participants in \cite{barke_grounded_2023} who say, e.g., \textit{“I don’t see the error immediately, and unfortunately, because this is generated, I don’t understand it as well as I feel like I would’ve if I had written it”}. Participants in \cite{weisz_better_2022} noted a trade-off between writing and debugging code, citing a lack of comprehension for AI-generated code translation and \textit{“spotting errors in ‘foreign’ code”} as challenges. Similarly, in data science, users report feeling out of control when unable to understand AI-generated suggestions \cite{mcnutt_design_2023} and highlight readability “as being a critical feature of usable synthesized code” \cite{drosos_wrex_2020}. For novices in a domain, this reduced situational awareness can be particularly challenging, as noted in \cite{prather_its_2023}. In the healthcare domain, AI-generated medical records may lead physicians to become detached from patients' medical history, and in turn spend additional time analysing GenAI outputs to compensate for the missing information \cite{preiksaitis_chatgpt_2023}. %
These findings indicate that gaining situational awareness of GenAI output is demanding and takes users’ time and attention away from proceeding with the main task. 

Automation research shows that low situational awareness can be exacerbated by several factors, including automation’s high output capacity and systems’ complexity, opaqueness, and low reliability.\footnote{Situational awareness can also be reduced due to automation-related \textit{unhelpful structuring of workflows} (Section \ref{subsec:restructuring}), including changes in the task sequence (Section \ref{subsubsec:sequence}) and the loss of feedback (see Section \ref{subsubsec:feedback}).} The next sections cover these factors, as well as a potential outcome of the `monitoring' challenge of automation: complacency and over-reliance.

\subsubsection{High automation capacity}\label{subsubsec:capacity}
Monitoring automation—in this case, evaluating GenAI output—is, ironically, made more difficult by the high capacity of automation, which makes it challenging to understand and anticipate system behaviour. For example, when traders in the digital stock exchange changed roles from executing to monitoring trades, they underperformed as they were unable to effectively monitor the trades in real-time \cite{haldane_systemic_2011}. As such, they resorted to monitoring them at a higher level of abstraction and required additional resources to process that information, thereby missing more trades that were executed in the meantime. 

Similarly, GenAI is notable for its high capacity in outputting content, such as entire documents or software programs, or multiple simultaneous suggestions \cite{barke_grounded_2023,schellaert_your_2023,chen_next_2023,sarkar_what_2022}. This makes evaluating these outputs challenging. In GenAI-assisted coding, \cite{barke_grounded_2023} found that users deal with the plethora of code suggestions by quickly assessing them using a “pattern matching” approach, where they search for the presence of certain keywords or control structures. The impact of high output capacity can be worsened by poor system design. For example, participants in \cite{barke_grounded_2023} noted that the separation of Copilot’s multi-suggestion pane from their main code increased cognitive load due to the lack of relevant code context when reviewing and trying to differentiate the code suggestions.   

\subsubsection{Automation complexity and opaqueness}\label{subsubsec:complexity}
Evaluation is further challenged by the complexity and opaqueness (i.e., poor explainability) of automated systems, which can reduce situational awareness. More features and modes create more possible interactions among system components and a corresponding reduction in system predictability as the system increasingly considers multiple factors or component states \cite{endsley_situation_2003}. This can lead to unfamiliar and infrequent system states, which add to the challenge of comprehending systems’ workings. For example, even well-trained pilots were startled by unexpected flight automation system behaviours in complex systems \cite{wiener_flight-deck_1980}. System opaqueness similarly reduces situational awareness and affects monitoring, for example, in the use of automation aids in local government organisations \cite{lindgren_ironies_2023}. Put another way, system complexity and opaqueness make it more difficult for users to create an accurate mental model of the system needed for the correct interpretation of information, including situations where manual control will be needed \cite{baxter_ironies_2012}.

The opaqueness and complexity of GenAI systems are cited as key barriers to usability, including prompting and evaluating outputs \cite{liao_designerly_2023, sun_investigating_2022}. One issue, termed `fuzzy abstraction matching' \cite{sarkar_what_2022}, describes the opaque relationship between the content of prompts and the resultant output, driven by the flexibility of GenAI models to produce plausible but potentially incorrect outputs for prompts with a wide range of abstraction. Another issue is the sheer range of implicit and explicit parameters available to users, which increases systems’ complexity \cite{schellaert_your_2023}. This not only makes prompting a challenge (e.g., \cite{zamfirescu-pereira_why_2023, dang_choice_2023}) but also the evaluation of outputs (e.g., \cite{weisz_perfection_2021, barke_grounded_2023, liang_understanding_2023}) as the two are inextricably intertwined in current systems. The top usability issue for AI programming assistants, as surveyed in \cite{liang_understanding_2023}, is not knowing what part of users’ code or comments the GenAI system is relying on to produce output. Likewise, one participant in \cite{barke_grounded_2023} laments the challenge of evaluating code suggestions, \textit{“it might be nice if it could highlight what it’s doing or which parts are different, just something that gives me clues as to why I should pick one over the other”}.

\subsubsection{Automation reliability}\label{subsubsec:reliability}
The challenge of monitoring automation is further exacerbated by systems’ unreliability. For example, \cite{metzger_automation_2005} found that air traffic controllers who worked with unreliable automation to make aircraft-to-aircraft conflict decisions were unable to monitor the systems effectively and were ultimately better at detecting conflicts without automation. Similar impacts of reliability were found for target detection and decision-making tasks \cite{galster_effects_2001, wickens_workload_2000}. Evaluation of GenAI outputs is likewise exacerbated by the non-determinism of GenAI models \cite{schellaert_your_2023}, which can produce different outputs for the same input, resulting in lower reliability from the user’s perspective. More than merely being non-deterministic, GenAI systems can introduce subtle or non-intuitive errors into outputs, particularly in long outputs such as multi-line code suggestions \cite{sarkar_what_2022} (see also Section \ref{subsec:polarization}). \citet{woodruff_how_2023} found that knowledge workers across domains overwhelmingly cited a lack of reliability as a key reason for humans having to review GenAI outputs. Example concerns ranged from violation of brand standards and copyrights in generated content, to inaccuracies in legal documents \cite{woodruff_how_2023}.

\subsubsection{Potential complacency and over-reliance}\label{subsubsec:overreliance}
Ultimately, as Human Factors research shows, the shift from production to evaluation, the resultant reduced situational awareness, and additional workload can result in complacency, over-reliance on systems, and increased errors \cite{parasuraman_humans_1997}. Trying to recover from these errors further increases the workload and, as workload affects monitoring ability, can create a vicious cycle. In high-workload situations, there are fewer attentional resources available for monitoring imperfect automation, resulting in a risk of errors \cite{mcbride_understanding_2011} and significantly longer error detection time \cite{dixon_mission_2005}. Complacency due to high-workload conditions has been observed in aviation, where pilots would fail to conduct sufficient checks of system state \cite{parasuraman_performance_1993, funk_flight_1999}. In a spacecraft simulator study, operators did not properly assess the recommendations and simply complied with them, which resulted in missed failures \cite{manzey_misuse_2006}. 

An increase in complacency and over-reliance related to output evaluation has been observed in GenAI user studies. For example, when verifying the correctness of AI-generated code, some programmers reported skimming through the output rather than reading and evaluating the code rigorously \cite{sarkar_what_2022, vaithilingam_expectation_2022}. This is especially prevalent for those with less experience, such as end-user programmers \cite{sarkar_what_2022} or novices \cite{prather_its_2023, kazemitabaar_how_2023}. In some cases, this has led to errors that users either missed \cite{ross_programmers_2023} or had to later spend time debugging \cite{vaithilingam_expectation_2022}. Notably, in advertising, both expert and non-expert writers showed overconfidence in the quality of AI-generated drafts, failing to thoroughly revise them \cite{chen_large_2023}. Complacency and over-reliance have also been reported in the data science domain \cite{gu_how_2023, gu_how_2023-1, srinivasa_ragavan_gridbook_2022}; in the legal domain, where "AI-assisted exams were more likely to miss hidden issues" \cite{choi_ai_2023}; and in the design domain, where one participant commented, \textit{“I would never design it like that, but this [GenAI system] thinks it can do it like that […] But this is what it gave me, so I don’t have a problem with that.”} \cite{gmeiner_exploring_2023}. Over-reliance has been shown to lead to decreased performance; for example, management consultants showed overall poorer performance when they blindly adopted AI-generated outputs \cite{dellacqua_navigating_2023}. 

\subsection{Unhelpful workflow restructuring}\label{subsec:restructuring}
Automation can restructure workflows in unhelpful ways by introducing new challenging tasks, disrupting familiar task sequences, and removing informative feedback (Figure \ref{fig:fig_challenges}b). This changes what strategies operators use, how they perceive information, and how they act in a specific context, potentially leading to ineffective use of freed-up time and cognitive resources. Thus, rather than reducing what they work on when all or part of tasks are automated, people instead rely on different strategies for working on that task \cite{bainbridge_ironies_1983}. For example, when automation introduces new tasks in operators’ workflow, disrupting their familiar workflow, they struggle to adapt their strategies \cite{klein_making_2006}. Likewise, when automation unexpectedly increases the workload during peak times, operators tailor the system or the task to accommodate the automation needs \cite{cork_development_1998}. If tailoring the system is not possible, users are forced to tailor their tasks, often having to add new tasks to their workload \cite{cork_development_1998}. For example, physicians using automation aids learned how to manipulate monitors displaying physiological data to fit their work strategies. However, because this manipulation was an additional task physicians had to perform, they avoided using the system in high-workload situations \cite{cork_development_1998}. Moreover, when automation changes the familiar sequence of the task, for example, by removing a step, operators make errors and repeat their actions. For example, physicians might forget to record a dose of medication in a log and mistakenly repeat the procedure \cite{altmann_brief_2015}. Finally, when automation removes the critical feedback necessary to make an informed decision, operators succumb to errors. For example, in aviation, pilots were missing critical failures due to relevant information from vibration and smell being lost in the automation process \cite{moray_acquisition_1986}.

\subsubsection{Prompting as a new task}\label{subsubsec:prompting}

The central role of prompting in GenAI systems is one major way in which such systems are restructuring workflows. Studies show that users struggle with prompting, dedicating considerable time and effort to it. In \cite{xu_-ide_2022}, programmers using a code generation plugin invested significant effort in experimenting with prompts to understand how their queries worked best. Likewise, in \cite{jiang_discovering_2022}, participants using an LLM-driven tool developed various strategies to deal with model failures, for example, rewording prompts by reducing the scope of the request or looking for alternative wording. Trying to adapt prompts is a cognitively demanding task, as participants must form a mental model of what the model can work with (the problem of `fuzzy abstraction matching'; \cite{sarkar_what_2022}). Beyond being demanding, prompting may interfere with other aspects of users’ workflows. For example, Copilot users’ code commenting workflows can change. Participants in \cite{barke_grounded_2023} wrote and re-wrote detailed comments intended for Copilot, hoping to increase the context available to the system, and then also spent time deleting comments for Copilot after the fact. 

Similar workflow changes were observed in the design and writing domains. For example, one non-professional designer in \cite{kulkarni_word_2023} complains, “it felt like I was fighting it…I felt like it was helpful, but I also felt like I had to massage every word and select every character very carefully not to upset it so that it could generate something I wanted” (see also \cite{oppenlaender_creativity_2022}). \citet{dang_choice_2023} distinguish between \textit{diegetic} prompts (instructions implicitly conveyed by inputted content to be acted on by the system) and \textit{non-diegetic} prompts (instructions explicitly conveyed to the system). The latter is particularly disruptive to users’ workflows in the writing domain, as they “[force] writers to shift from thinking about their narrative or argument to thinking about instructions to the system” \cite{dang_choice_2023} (see also \cite{yuan_wordcraft_2022}), a finding echoed in the coding domain \cite{jayagopal_exploring_2022}. %
More broadly, prompting seems to function as a new task that competes with other workflow tasks, adding to the workload and potentially increasing over-reliance on automation as users invest more time into it \cite{endsley_distribution_2016}. Indeed, this might explain why some users try to coerce AI output to be useful (see Section \ref{subsubsec:adaptation}) or become complacent in reviewing it (see Section \ref{subsubsec:overreliance}).      

\subsubsection{Output adaptation as a new task}\label{subsubsec:adaptation}

Another workflow change with GenAI is the need to adapt generated output, effectively a new type of task. In \cite{barke_grounded_2023}, several participants chose to adapt Copilot suggestions to use as a template for their code. Rather than accepting or rejecting code entirely, they deleted and edited parts so they would not have to write it from scratch. Others used the strategy of slowly breaking down large blocks of code and adapting them as needed or cherry-picking code from multiple suggestions. This suggests that the use of suggestions is not straightforward, and complex strategies are created by programmers for their workflows. The productivity gains of these workflow changes remain unknown, and although participants in \cite{barke_grounded_2023} found them helpful, they may ultimately decrease productivity. For example, if the adapted code has an error, the necessary debugging will add to the workload, as observed in, e.g., \cite{barke_grounded_2023} and \cite{vaithilingam_expectation_2022}. In the design domain, \cite{gmeiner_exploring_2023} found that manufacturing designers struggled with GenAI assistance. In this case, the GenAI system was found to be “dominating the design process”, and “designers either gave up and accepted unsatisfying results, improvised ‘hacky’ strategies to work around the AI or abandoned the AI assistance altogether and proceeded to work manually”.

The productivity gains or losses of output adaptation may depend on users’ expertise. %
In \cite{vaithilingam_expectation_2022}, participants of varying levels of expertise struggled to adapt the code suggestions, and many abandoned them entirely, thereby losing time. Among novices, code adaptation may particularly reduce productivity. \citet{prather_its_2023} studied novice programmers working with Copilot, identifying an unproductive interaction mode they termed “shepherding”, in which participants spent considerable time trying to coerce Copilot to produce useful code. This included accepting suggestions, then deleting them without any changes, or spending considerable time adapting suggestions without writing any code of their own. More broadly, the assortment of code adaptation strategies reflects a new layer of complex tasks that programmers are introducing to their workflow to accommodate and effectively use GenAI. Ironically, the more complex the code, the more powerful the potential productivity benefits, yet the more intricate and time-consuming the process of reviewing and adaptation might become (e.g., \cite{barke_grounded_2023}).

\subsubsection{Loss of task sequence}\label{subsubsec:sequence}
Workflow changes can also lead to difficulty in following the familiar sequence of steps in a task. Many tasks have sequential constraints, a set of steps that have to be performed in a specific order. When one of the steps is skipped or repeated, errors can occur \cite{altmann_brief_2015}. To perform a task correctly under sequential constraints, the cognitive system has to keep track of where it is in the sequence and select the correct next step when one step is complete \cite{altmann_brief_2015}. Changes in the structure of the task can make it difficult for one to follow the natural sequence of the steps. Automation research showed that operators' reactions are slower and less integrated when they cannot generate the sequence of activity themselves \cite{janssen_integrating_2015}. Not having a task structure to follow also prevents users from monitoring their own progress. Under manual control, users obtain information about the results of their actions and then can correct themselves \cite{smith_simulator_1979}. Without this information, they are more likely to repeat the same type of errors \cite{wiener_flight-deck_1980}. 

In GenAI workflows, auto-suggestions generated by the system or the requirement to prompt systems are examples of disruptions to the familiar sequence of steps, which could lead to productivity loss, as evidenced in recent studies. In the coding domain, \cite{barke_grounded_2023} found that long code suggestions in Copilot disrupted users’ task sequence by “forcing them to jump in to write code before coming up with a high-level architectural design”. Analogously, in the design domain, \cite{gmeiner_exploring_2023} found that the need for prompting meant that designers had to specify required parameters in advance instead of working step-by-step, thereby requiring designers “to think through the design problem in advance, which is challenging and different from the usual iterative design process”. This loss of task sequence can be particularly disruptive among novices. For example, \cite{prather_its_2023} identified an unproductive interaction pattern among novice programmers called “drifting”, in which participants spent time adapting code suggestions, then deleting them, and repeating the cycle. Thus, they unproductively drifted from suggestion to suggestion without a direction. Moreover, this was exacerbated if the generated output contained an error, which sent users down a “debugging rabbithole”, in which they spent time trying to adapt incorrect code rather than focusing on the correct solution \cite{prather_its_2023}. In film production, \citet{parra_pennefather_ai_2023} observed a filmmaker working with GenAI that had to shift between multiple software, struggling to identify which was the most suitable for which part of their creative process. The creative described the process as \textit{"an exercise in randomization and an attempt to control chaos"} \cite{parra_pennefather_use_2023} (see \cite{oppenlaender_creativity_2022} for similar observations with creative text-to-image generation workflows).

Task sequence can also be obscured when a large part of the workflow is automated. For example, both expert and non-expert copywriters were anchored to GenAI suggestions and produced lower-quality results when GenAI generated the majority of the text versus when it only provided feedback to users \cite{chen_large_2023}. Similarly, professional novel writers \cite{calderwood_how_2020} and inexperienced writers working with GenAI \cite{arnold_generative_2021} found %
guidance more useful than the injection of %
generated text.%
In these examples, users’ familiar task sequences in a given domain are disrupted by aspects of GenAI systems.

\subsubsection{Loss of feedback}\label{subsubsec:feedback}
Automation can deprive users of key feedback needed to assess the state of automation and its ability to perform tasks. For example, automation can cause users to change from processing raw data to processing integrated information. Introducing automation into paper-making plants moved operators away from the information associated with informal feedback (e.g., smells, sounds) and put them in control rooms \cite{lee_human_2009}. This change not only required operators to learn the task of plant control but also deprived them of contextual information that could help them diagnose automation failures and intervene appropriately. Similarly, in aviation, relevant information from vibration and smell was lost in the automation of process control operations \cite{moray_acquisition_1986}, and the automation of auto-feathering systems in commercial aircraft removed the signal telling pilots about engine shut-downs \cite{billings_toward_1991}. The lack of transparency or supporting contextual feedback often only becomes an issue under system failures when operators lack the relevant detail for detecting or addressing them \cite{endsley_effect_1997}.

An analogous loss of feedback has also been observed in GenAI-assisted coding. Participants in \cite{vaithilingam_expectation_2022} noted that, in comparison to internet search tools like Stack Overflow, Copilot lacked additional information, such as discussions, explanations, and comparisons of code solutions. This sentiment was echoed by participants in \cite{ross_programmers_2023}, who noted that their AI code assistant “lacked the ‘multiple answers’…and ‘rich social commentary’…that accompanies answers on Q\&A sites”. Thus, programmers using these tools see the code, comments, and data but miss out on the rich feedback that is usually available when programming with access to various media sources. 

\subsection{Task interruptions}\label{subsec:interruptions}
Another aspect stifling productivity gains from GenAI is task interruption (Figure \ref{fig:fig_challenges}c). There are various cognitive costs related to interruptions \cite{altmann_memory_2002, janssen_identifying_2011, salvucci_toward_2011}. Interruptions can disrupt the user’s thought processes \cite{altmann_momentary_2014} and initiate a switch between tasks that requires time and cognitive resources, which negatively affects performance \cite{janssen_integrating_2015}. Particularly long and complex interruptions significantly disrupt people’s ability to resume their original tasks \cite{mark_pace_2012, monk_effect_2008, mark_cost_2008}. Moreover, interruptions can also break the user’s flow state \cite{taekman_virtual_2010}. 

Copilot auto-suggestions have been shown to interrupt users’ main tasks, with programmers referring to Copilot auto-suggestions as \textit{“interrupting their thoughts”} \cite{sarkar_what_2022}, \textit{“intrusive”}, and \textit{“messing up thought process”} \cite{prather_its_2023}. Accordingly, some programmers decide to switch the suggestions off to avoid distractions \cite{sarkar_what_2022} or chose to disable the tool completely \cite{barke_grounded_2023}, while others admitted being \textit{“tempted to follow what it’s saying instead of just thinking about it”} \cite{prather_its_2023}. Beyond programming, similar interruptions are reported in the writing domain \cite{clark_creative_2018,dang_choice_2023,bhat_interacting_2023} and in data science \cite{mcnutt_design_2023,gu_how_2023-1}.

Particularly distracting are the long, multi-line code suggestions. For example, these have been observed to break programmers’ flow when in `acceleration mode’, a state in which programmers work with well-formed intent, relative to an `exploration mode’, in which programmers start a novel task or debug \cite{barke_grounded_2023}. Programmers were distracted from their flow as they felt compelled to read the code. If they chose to consider it, they then had to review it for errors. Thus, long code suggestions force users to switch back and forth between writing and reviewing code, and if the code has errors, they must then switch to debugging \cite{vaithilingam_expectation_2022}. This may be particularly disruptive if the errors are unrelated to the current task focus, as found in \cite{weisz_perfection_2021}. Interruptions may be particularly impactful for novice programmers, who are tempted to read the large blocks of code despite their perception as a nuisance \cite{prather_its_2023}. Accordingly, their attention is shifted from thinking and problem-solving to deciphering. Ironically, the feature that should accelerate productivity significantly increases participants’ cognitive load due to the associated task-switching. 

Programmers, particularly experienced ones, eventually learned to dismiss long, multi-line suggestions \cite{barke_grounded_2023,sarkar_what_2022}. Nevertheless, even when ultimately rejecting these, their thought processes were already disrupted. This was the case not only for novice programmers who reported \textit{“[wasting] time reading instead of thinking”} \cite{prather_its_2023}, but also for experienced programmers: \textit{“I was about to write the code, and I knew what I wanted to write. But now I’m sitting here, seeing if somehow Copilot came up with something better than the person who’s been writing Haskell for five years…”} \cite{barke_grounded_2023}. Similarly, in the writing domain, some users learned to ignore suggestions in certain contexts, whereas others deliberately sped up their writing to avoid getting distracted by a suggestion \cite{bhat_interacting_2023}. 

That complex code suggestions are the most distracting during `acceleration’ and are more helpful during `exploration’ \cite{barke_grounded_2023} suggests that their timing is a key factor. Indeed, automation research speaks to this. People respond faster to interrupting tasks if the interruption was scheduled as a breakpoint between main task chunks \cite{iqbal_effects_2008} or when they occur at subtask boundaries \cite{bailey_understanding_2008, iqbal_investigating_2005, janssen_strategic_2010}. Similarly, \cite{cutrell_effects_2000} found that users interrupted earlier in a task were more likely to request a reminder after being interrupted, and \cite{cutrell_what_2007} showed that the later in the main task an interruption occurs, the less recovery time is needed when subsequently returning to it. Indeed, in the data science domain, \cite{gu_how_2023} found that when AI suggestions were out of sync with users’ current analysis plans, participants were either distracted or ignored them.

\subsection{Task-complexity polarization}\label{subsec:polarization}

Automation often makes easy tasks easier but fails to reduce the workload during cognitively demanding tasks, and in fact, often makes them harder \cite{lee_human_2009}. This has been termed “clumsy automation” in Human Factors research \cite{cook_cognitive_1991}, but we introduce the more precise term \textit{task-complexity polarization} (Figure \ref{fig:fig_challenges}d). One explanation is that easy tasks are easier to automate, and so the more difficult tasks tend to remain under manual control, albeit alongside the additional task of monitoring automation, and within a now more fragmented workflow \cite{lee_human_2009}. For example, automation has been shown to reduce pilots’ mental workload when it is already low during easy tasks, as when the plane is on autopilot during a straight flight. However, automation increased the mental workload of pilots when the flight-related workload was already high, e.g., during landing, as they then had to simultaneously reprogram the system managing autopilot, activate landing procedures, and manage communication \cite{wiener_flight-deck_1980}. Humans are also ineffective in shifting cognitive resources saved by automation to support more difficult tasks. In the study by \cite{metzger_automation_2005}, air traffic controllers used automation designed to aid conflict detection and resolution tasks. This was expected to free up enough mental resources that controllers could allocate to performing more complex tasks. However, automation did not reduce the mental workload in routine tasks that were demanding, such as communication and accepting and handing off aircraft. Either the aid did not free enough resources, or the controllers could not allocate them to improve communication performance. Studies on automated decision-making used to support government tasks showed that the new technology often only reduced the easy assignments but left the difficult ones to the government workers, making their work more difficult and fragmented \cite{lindgren_ironies_2023}. 

GenAI studies show that a similar pattern is emerging in current users of GenAI systems. First, there is evidence that GenAI systems are most helpful at making easy tasks even easier. For example, GenAI has been shown to be the most effective in supporting novice writers performing easy assignments and low-skilled customer service agents in entry-level tasks \cite{frey_generative_2023}. In AI-assisted programming, users across studies were most confident in using GenAI for simpler tasks, such as “writing boilerplate, repetitive code” \cite{barke_grounded_2023}, “short chunks of code” \cite{ross_programmers_2023}, or “coding in narrow contexts” \cite{sarkar_what_2022}. Barke et al. \cite{barke_grounded_2023} found that the most successful Copilot users were able to decompose the coding task into “microtasks”, which Copilot was effective at completing (see also \cite{vaithilingam_expectation_2022}. However, it is precisely the task decomposition process itself that is the more cognitively demanding task, and for which Copilot was not able to provide support. Indeed, Copilot’s limitations with larger coding problems meant that “[it] led to more task failures in medium and hard tasks” \cite{vaithilingam_expectation_2022} (see also \cite{ross_programmers_2023, sarkar_what_2022}). In the data science domain, some users similarly reported feeling most confident in relying on GenAI for “peripheral tasks such as error-checking or report generation, rather than the central analysis process” \cite{gu_how_2023}. Likewise, in a study of AI-assisted legal analysis using GPT-4, Choi and Schwarcz \cite{choi_ai_2023} conclude that "AI helps with simple legal analysis but stumbles over complex legal reasoning". Thus, whereas GenAI succeeds at making easy tasks even easier, current systems are less effective at supporting harder tasks. 

There is also evidence that GenAI can make hard tasks even harder. First, as discussed throughout, GenAI systems can shift users’ roles to one of cognitively demanding output evaluation (Section \ref{subsec:shifteval}), restructure workflows in unhelpful ways (Section \ref{subsec:restructuring}), and interrupt workflows (Section \ref{subsec:interruptions}), all of which can interfere with users as they work on demanding tasks, for example by depriving them of relevant context or disrupting their task sequence. This can be particularly disruptive for novices, as one participant noted about long code suggestions, \textit{“if you do not know what you’re doing, it can confuse you more”} \cite{prather_its_2023}. 

Secondly, GenAI systems can introduce errors into outputs that users must deal with. In AI-assisted coding, GenAI systems can “introduce subtle, difficult-to-detect bugs, which are not the kind that would be introduced by a human programmer writing code manually” \cite{sarkar_what_2022}. Errors are particularly likely in longer code suggestions \cite{barke_grounded_2023, sarkar_what_2022}, precisely the ones that might help users address complex tasks. This makes the already demanding task of debugging even more difficult, not only because of the inherent challenge of debugging `foreign’ code (as discussed in Section \ref{subsec:shifteval}), but also because of errors’ subtlety and the difficulty in discerning whether an error is the user’s or the system’s fault \cite{barke_grounded_2023, vaithilingam_expectation_2022, sarkar_what_2022}. A similar concern about GenAI systems introducing errors has been raised in the data science domain \cite{gu_how_2023}.   

Thirdly, when users are stuck on a demanding task, although GenAI systems can provide multiple suggestions to help, this ends up overwhelming some users. Weisz et al. \cite{weisz_better_2022} found that users’ frustration and mental demand were significantly heightened when multiple AI-generated code translations were shown to participants. Users similarly found the multi-suggestion pane in Copilot to be overwhelming when they accessed it during a state of coding “exploration” (i.e., starting a novel task or stuck on a task \cite{barke_grounded_2023}). Thus, ironically, GenAI systems can make hard tasks even harder in various ways that may ultimately leave users with the same or %
increased cognitive workload.    

\section{Human Factors Solutions}\label{sec:solutions}
Beyond diagnosing the usability challenges of automation, Human Factors research has spent decades studying approaches to mitigate these challenges (e.g., \cite{endsley_here_2017, sheridan_human-automation_2005, parasuraman_model_2000, parasuraman_effects_1997}). Here, we outline some key potential design solutions that could reduce the productivity loss in human-GenAI interaction. These include providing continuous relevant feedback to users (Section \ref{subsec:contfeedback}), enabling system personalization (Section \ref{subsec:pers}), applying ecological interface design (Section \ref{subsec:eid}), using task stabilization and interruption timing techniques (Section \ref{subsec:taskstab}), and enabling clear task allocation between users and systems (Section \ref{subsec:taskalloc}). Besides targeting individual productivity loss challenges, these solutions share the underlying Human Factors principles of providing feedback and enabling system flexibility \cite{carayon_human_2019}.  

More broadly, we argue that these proposed approaches aim to (i) increase user agency in how they adapt the GenAI support to users' preferred ways of working, reducing the cognitive load stemming from disrupted workflows; (ii) increase users' situational awareness of system changes and potential errors, reducing the cognitive load associated with the monitoring and evaluation of AI outputs; and (iii) increase user flexibility through the more granular application of AI support to their tasks, freeing users from having to make a binary decision of either using GenAI tools potentially ineffectively or not using them at all \cite{sarkar_what_2022, chen_next_2023}. Throughout, we focus on the programming domain as an example of %
how these approaches can be applied to GenAI systems. 

\subsection{Continuous feedback}\label{subsec:contfeedback}
When GenAI is introduced to users' workflows, their role can shift from active involvement in performing the task (i.e., production) to more passively reviewing the AI-generated outputs for errors (i.e., output evaluation). The latter is a cognitively demanding task due to the lack of supporting contextual information and the resultant loss of situational awareness. We propose that feedback about system behavior is a key strategy to keep users engaged and in the loop of GenAI system performance. 

During the monitoring stages, receiving continuous feedback is crucial for the operator to remain in the loop and recognise moments when interruption and input are needed \cite{loft_modeling_2007, lee_human_2009}. Feedback is essential to help operators know if their requests have been received if the actions of the automation system are being performed properly, and if any errors are occurring \cite{norman_problem_1997}. With GenAI systems, this includes knowing which aspects of the input are serving as prompts, how they are being interpreted by the system, how the output matches them, and whether there are any errors. Thus, feedback is tied to carefully designed explainability features \cite{liao_ai_2023,sun_investigating_2022}. It should help users know why the system is responding in a certain way and allow them to build mental models of the system's behaviour, how it interacts with them, and where they can expect failures (i.e., cause-and-effect relationships). Moreover, by helping users develop a more accurate mental model of the system, feedback can also serve to support users in better prompting and output adaptation \cite{chen_next_2023,liao_ai_2023}, thereby helping them structure their workflows more effectively. 

We suggest that GenAI tools should continuously provide relevant feedback to users, updating them on the system's state, particularly during the monitoring stages. Feedback should be informative but non-intrusive, where the amount and form of feedback adapts to the interactive style of the participants and the nature of the problem \cite{norman_problem_1997}. 

In the context of GenAI systems, feedback is important for understanding system inputs and outputs and the cause-and-effect relationship between them. In the case of automated suggestions, users expressed a need for more information on specifically which code and comments Copilot relies on as inputs \cite{barke_grounded_2023}. In the case of conversational interfaces, feedback could highlight prompt changes and the resulting output changes \cite{zamfirescu-pereira_why_2023}. Feedback could also be used to support pattern-matching between the AI suggestions and users' task goals. For example, the output could have keywords highlighted, such as function calls or variable names, that would be a meaningful indication of a code fit \cite{barke_grounded_2023}. It could also include more context and documentation with the output, e.g., links to Stack Overflow or official documentation pages \cite{xu_-ide_2022}, or provide relevant usage examples \cite{moreno_how_2015}. To understand outputs, \citet{vaithilingam_expectation_2022} suggested using inline comments or highlighting different parts of the code based on confidence to help users understand the code generated by Copilot (see also \cite{weisz_perfection_2021, vasconcelos_generation_2023}). The authors also suggested supporting debugging by automatically generating test cases and test data for users to validate and identify corner cases \cite{vaithilingam_expectation_2022}. \citet{weisz_better_2022} proposed using alternate translations, where the system showed users the alternative it had considered to help them identify errors. In the writing domain, \cite{yuan_wordcraft_2022} proposed that systems should give prompt suggestions to users. 

Feedback can be overwhelming if it is poorly presented or excessive. It can also be incomprehensible without proper context, abstraction, and integration \cite{lee_human_2009}. As such, feedback should be provided by applying methods of ecological interface design \cite{rasmussen_coping_1989} (see Section \ref{subsec:eid}) and notification design \cite{paul_interruptive_2015} (see Section \ref{subsec:taskstab}), which are effective approaches for improving situational awareness and error detection.

\subsection{System personalization}\label{subsec:pers}
Human Factors studies have shown that when system personalization is constrained, the cognitive demands on operators and the associated productivity loss both increase \cite{cook_adapting_1997}. Indeed, as described in Section \ref{subsec:restructuring}, increased cognitive demand and productivity loss have been observed in studies of GenAI-assisted programming as users try to understand and accommodate systems by changing their ways of working. This could be mitigated by allowing users to flexibly personalize systems to fit their tasks and ways of working \cite{lee_human_2009}. 

For example, users could personalize the system to provide help when needed rather than having suggestions generated automatically. In creative writing, choosing when to receive feedback from GenAI, rather than receiving AI-generated text, preserved writers' creativity and alleviated anchoring effects and over-reliance \cite{chen_large_2023}. Users should also be able to inform the system about their state of work (e.g., `acceleration' or `exploration', as per \cite{barke_grounded_2023}), so suggestions would better match the users' goals in terms of complexity, variety, length, and frequency \cite{gu_how_2023-1}. Systems could automatically detect users' states \cite{barke_grounded_2023, gu_how_2023-1}, guided by user-adjustable parameters, and respond according to user-provided preferences \cite{rao_can_2023}, feedback \cite{madaan_language_2022}, or through the use of prompts \cite{wu_personalized_2022}. Users should also be able to personalize the inputs to the system. For example, \cite{barke_grounded_2023} proposed that users should be able to control the context they provide to Copilot, enable comments that make code invisible to the tool, or decide that the tool will rely on Stack Overflow-style prompts rather than in-context code. 

Personalization is particularly important as users might have varying levels of task and domain expertise, which has been shown to affect their preferences and needs regarding the amount and kind of information provided \cite{paris_tailoring_1988}. For example, novice programmers might want to spend some time working on the problem themselves and only ask Copilot for support when they are stuck \cite{prather_its_2023}, whereas experts might want to simply complete their lines \cite{barke_grounded_2023}. 

\subsection{Ecological interface design}\label{subsec:eid}
The introduction of GenAI to users' workflows can disrupt them, leaving workers looking to adjust their ways of working or their familiar task structure. These processes increase cognitive load and result in productivity losses. Moreover, these disruptive changes can prevent users from being able to exercise their expertise and from benefiting from AI support. To align GenAI systems with users' workflows effectively, we suggest that GenAI systems be designed according to an ecological interface design (EID) approach. EID emphasizes designing interfaces that reflect users' perceptual constraints within a work environment in a highly domain- and context-specific manner \cite{rasmussen_coping_1989}. Specifically, it emphasizes (i) combining what users control and what they see in the system so that they can interact using clear, real-time signals; (ii) providing a consistent mapping between work domain constraints and interface cues; and (iii) showing the system's key relationships directly on the screen, making it easier for users to form a mental model of the system \cite{rasmussen_coping_1989,mcilroy_ecological_2015}. EID has been shown to reduce workload and improve performance in aviation risk management \cite{king_clear_2022}, medical domains \cite{effken_making_1997}, and automation-assisted driving \cite{stoner_applying_2003}. 

In practice, this approach suggests that an automation aid or AI system should be designed to perform consistently with operators’ mental models, preferences, and expectations in a given work domain \cite{goodrich_seven_2003}. For example, GenAI systems should consider a broader domain context for their inputs by including information from interactions with external sources within the work domain (e.g., with Copilot, the consideration of code beyond the current file \cite{bird_taking_2023}). Which sources and when they are considered should be clearly specified to users to support real-time control. 

Systems should also consider work domain constraints. For example, Copilot should consider the natural task sequence of certain programming tasks by providing support for high-level architectural design (or planning) when it is needed and avoiding code suggestions that might interfere with this process \cite{gu_how_2023-1} (see also Section \ref{subsec:taskstab} for more on managing interruptions). Likewise, interfaces should adapt to support debugging when long code suggestions are provided, as outlined in Section \ref{subsec:contfeedback}. Systems should also help users understand how code suggestions map to and affect other aspects of the code beyond the local insertion point. Likewise, when helping physicians with administrative tasks, GenAI system outputs should include records of patients' unique medical histories and physicians' clinical reasoning \cite{preiksaitis_chatgpt_2023}. 

EID also aims to support users’ ways of perceiving information in a specific domain. For example, it encourages using a hierarchical visual structure to display relevant information to allow multiple levels of information to be (meaningfully) visible simultaneously in the interface. This way, users can guide their attention to the level of interest, depending on their level of expertise and current task demands \cite{rasmussen_coping_1989}. This also supports flexibility, as users do not have to attend to a specific description level. For example, depending on where users are in their workflow, GenAI systems can provide programmers with suggestions at different levels of abstraction \cite{gu_how_2023-1}, from high-level pseudo-code to low-level implementations, organized in a visual hierarchy, which would be particularly helpful for novices \cite{prather_its_2023}. Similarly, \cite{gu_how_2023-1} suggest that interactive visualizations, linked to users' code and other parts of the interface, can be used to support decision-making. 

Finally, as discussed in Section \ref{subsec:contfeedback}, explainability features are essential to help users form an accurate mental model. These features should be integrated directly into the interface (e.g., as in AI Chains \cite{wu_ai_2022}), taking into consideration the work domain context. In the healthcare domain, explainability has been shown to be most effective when combined with insights from medical experts. Without considering domain specifics, explanations lacked important context and included unnecessary information (e.g., background skin texture) that confused expert dermatologists \cite{degrave_dissection_2023} (see also \cite{huang_generative_2023} for similar results in radiology). 

\subsection{Main task stabilization and interruption timing}\label{subsec:taskstab}
As discussed in Section \ref{subsec:interruptions}, GenAI system suggestions (e.g., Copilot code suggestions) interrupt users, especially during their flow states, distracting them and potentially leading to productivity loss. Accordingly, some users disable auto-suggestion features or GenAI systems entirely because of their distracting nature \cite{barke_grounded_2023, sarkar_what_2022,liang_understanding_2023}. Writers similarly prefer not to be interrupted by AI-generated snippets of text \cite{chen_large_2023}. Instead of forcing users to avoid interruptions by disabling tools, systems should preserve users’ flow states by incorporating task stabilization techniques or by carefully timing interruptions around their flow states. 

\subsubsection{Task stabilization via attention guidance}\label{subsubsec:stabatt}
Interruptions can be designed to support task stabilization, i.e., to help users prepare their current (main) task for the temporary switch in focus \cite{czerwinski_diary_2004, parnin_evaluating_2010}. For example, among software users and developers, \cite{paul_interruptive_2015} found that interruptions were helpful when they directed users to the parts of the current task (or a new task) they needed to attend to. Interruptive notifications were also useful as progress indicators, helping users plan and resume their next task after interruption. In the case of GenAI systems such as Copilot, this could manifest in long code suggestions being divided (e.g., via colour) into small logical units for programmers to easily parse during the acceleration (flow) mode \cite{barke_grounded_2023}. Alternatively, systems could direct users’ attention to certain keywords (e.g., via highlighting) that could help them identify the applicability of the suggestion by using “pattern matching” \cite{barke_grounded_2023}. In line with Human Factors principles, interface design should provide cues to guide users’ attention to the next appropriate action. Otherwise, users may fall into `procedural traps’ \cite{rasmussen_coping_1989, reason_contribution_1997}, novel situations where they rely on their normal rule set but without the usual success. Indeed, this has been observed in Copilot studies, where programmers end up in `debugging rabbitholes’ \cite{prather_its_2023, vaithilingam_expectation_2022}.

\subsubsection{Task stabilization via pre-interruption alerts}\label{subsubsec:stabpre}
Task stabilization can also be achieved by using pre-interruption alerts, which function as progress indicators, helping users plan and resume their next task after interruption \cite{paul_interruptive_2015}. Andrews et al. \cite{andrew_humans_2003} found that those who received a pre-interruption alert could resume the main task faster than participants who did not. This aligns with studies showing that adding a brief lag period before interruption helps users set place-keepers at their current task point, making it easier for them to return to it after being interrupted \cite{altmann_brief_2015, brumby_recovering_2013}. Similar pre-interruption alerts may be helpful for GenAI systems. For example, when Copilot is about to suggest a long code chunk, an alerting notification could create a brief pause period necessary to lock the users’ main task state. Even better, AI systems should set place-keepers automatically together with auto-suggestions, along with any other context-relevant information that could help users return to their train of thought. This would begin to address the challenge of helping users regain their prior context post-interruption, as has been raised in GenAI-assisted coding \cite{ross_programmers_2023} and data science \cite{gu_how_2023-1, gu_how_2023}. 

\subsubsection{Timing of interruptions}\label{subsubsec:timing}
Timing interruptions thoughtfully is another way to reduce their associated productivity loss. Interruptions are valuable for user productivity when they provide valuable awareness about things outside the user’s attention, such as new or background tasks \cite{paul_interruptive_2015}. However, interruptions can be disruptive when related to a task currently in focus. We propose that systems such as Copilot should be able to recognise when the user is in focus \cite{barke_grounded_2023, gu_how_2023-1}. Then, interruptions should be limited to supporting contextual alerts or providing information about ongoing tasks in the background (e.g., providing explainability information). Otherwise, during this stage, suggestions should carefully align with users’ flow \cite{gu_how_2023-1}, in line with ecological interface design. The system should recognise the strategies that users use during the flow state and support them by completing their thought processes, for example, auto-completing the end of the code line \cite{barke_grounded_2023}, providing only short code suggestions \cite{prather_its_2023}. Recognising when users are not in a flow state, systems could give users prompt examples and suggestions \cite{yuan_wordcraft_2022}, provide feedback \cite{chen_large_2023}, or goal-orientated guidance \cite{arnold_generative_2021}. This could be supported further by user personalization as per Section \ref{subsec:pers}. This would enable GenAI support to be used more narrowly (e.g., to provide warning messages and supporting contextual information or short snippets of code) rather than users having to use the GenAI ineffectively or turn it off completely. 

\subsection{Clear task allocation}\label{subsec:taskalloc}

GenAI user studies suggest that current systems make easy tasks easier and hard tasks harder for users, a phenomenon we have termed task-complexity polarization (and referred to as "clumsy automation" in the Human Factors literature \cite{wiener_flight-deck_1980}). Thus, it appears that these systems are not applied effectively to reduce overall workload. Human Factors research shows that one of the ways to address this is by clearly specifying how tasks are allocated between the human and system, particularly during high workload periods \cite{enstrom_real-time_1977,wallace_sinaiko_human_1972}. This not only better distributes the workload according to the respective strengths and weaknesses of humans and automated systems but also reduces the cognitive demand on users stemming from trying to discern the relative responsibilities on a moment-by-moment basis. For example, in aviation, reducing pilots' workflow to a single loop (eliminating the need for the operator to interact with the automation through the high workload tasks) resulted in better performance in a cockpit simulator. Similarly, allocating tasks to the computer and allowing the operator to deal with the queue items manually have also been shown to reduce workload \cite{chu_adaptive_1979}. We suggest that the allocation of tasks between the user and GenAI system should be clearly defined and supported by GenAI systems. The user should know which tasks the GenAI system deals with at a given moment. \cite{cook_adapting_1997}

As discussed in Sections \ref{subsec:shifteval} and \ref{subsec:polarization}, for simple tasks or in low workload conditions, users often let the GenAI system operate continuously. However, when complex tasks needed to be performed, they often stepped in and overrode the system and, in some cases, engaged in ineffective practices (e.g., reviewing code suggestions, editing, and then deleting them \cite{prather_its_2023}). Instead of having to do this, users should be able to proactively allocate responsibilities to the GenAI system. For example, according to their experience with the system, personal preferences, or expertise, they could identify tasks or parts of the tasks that they are confident that AI will perform successfully without their oversight or ones that they found AI to be most helpful with. For example, users might prefer manually translating certain types of code \cite{weisz_better_2022} or allowing the tool to be responsible for generating control structures while the user fills out the body \cite{barke_grounded_2023}. Likewise, users could allocate only repetitive `boilerplate' code for the system to complete autonomously while requesting its high-level planning support (rather than entire code completion) during more complex or exploratory tasks. In creative domains, this might mean that GenAI tools provide ideas in an open-ended form (e.g., probing questions), rather than as explicit suggestions \cite{arnold_generative_2021}, an approach that was found to be particularly helpful in copywriting \cite{chen_large_2023}. Making this initial allocation of responsibility and clearly understanding how tasks are divided would reduce the cognitive load of interacting with the GenAI system throughout demanding tasks. Moreover, it would help users better manage their demanding role as evaluators of AI output (as per Section \ref{subsec:shifteval}). 

Supporting effective task allocation depends on GenAI systems having a clear understanding of the work domain context, which is enabled by ecological interface design (see Section \ref{subsec:eid}). As such, the described Human Factors approaches work in synergy to support human-GenAI interaction and productivity.

\section{Conclusions}\label{sec:conclusions}

We have synthesized and analyzed the productivity challenges emerging during human-GenAI interactions, focusing on the much-studied domain of software development and noting similarities in areas such as data science, design, and writing. We have demonstrated the parallels between productivity challenges in older Human Factors automation studies and recent GenAI studies. Drawing on the human automation studies, we %
have categorised these challenges and the underlying reasons related to Human Factors, such as workload, feedback, and situational awareness. We show how aspects like the shift from active production (e.g., writing code) to passive evaluation (e.g., reviewing code), unhelpful workflow restructuring, task interruptions, and task-complexity polarization can stifle human performance and effective implementation of GenAI. 

Further extrapolating from human-automation studies, we have provided a set of design solutions that could help avoid productivity losses in human-GenAI interaction. More broadly, we argue for more consideration of users’ workflows, unique ways of working, and domain specificities when designing GenAI tools. To achieve this, we propose that systems be designed in accordance with ecological interface design, the principle of continuous feedback, support for flexibility via task allocation between users and systems, and user-guided system personalization. We also provide concrete design solutions for effectively guiding user attention during interruptions. 

Our paper is an initial bridge between Human Factors and Human-Computer Interaction issues of human-GenAI interaction. There is, of course, far more nuanced Human Factors research that can help understand and address the key productivity challenges in this fast-paced area. %
Reciprocally, we also expect that future Human-Computer Interaction research may open up new domains of exploration for Human Factors. 

\begin{acks}
Anonymized.
\end{acks}

\bibliographystyle{ACM-Reference-Format}
\bibliography{ProductivityLoss_refs}

\end{document}